\documentclass[preprint,hyper]{JHEP3} 

\JHEPspecialurl{http://jhep.sissa.it/JOURNAL/JHEP3.tar.gz}

\usepackage{epsfig,multicol}

\newcommand{\beq}{\begin{equation}}

\newcommand{\eeq}{\end{equation}}
\newcommand{\bea}{\begin{eqnarray}}
\newcommand{\eea}{\end{eqnarray}}
\newcommand{\rf}[1]{(\ref{#1})}

\newcommand\fverb{\setbox\pippobox=\hbox\bgroup\verb}
\newcommand\fverbdo{\egroup\medskip\noindent%
                        \fbox{\unhbox\pippobox}\ }
\newcommand\fverbit{\egroup\item[\fbox{\unhbox\pippobox}]}
\newbox\pippobox

\title{Integrable Spin Chains with $\mathop{\rm U}(1)^3$ symmetry
and generalized Lunin-Maldacena backgrounds}

\author{L. Freyhult, C.\ Kristjansen and T. M\aa{}nsson\\
        NORDITA, Blegdamsvej 17, DK-2100 Copenhagen\\
        E-mail: \email{freyhult@nordita.dk}, \email{kristjan@nbi.dk}, \email{teresia@nordita.dk}}

\preprint{\hepth{0510221}\\NORDITA-2005-68}     
\abstract{
We consider the most general three-state spin chain with $U(1)^3$ symmetry
and nearest neighbour interaction. 
Our model contains as a 
special case the spin chain describing the holomorphic three scalar sector
of the three parameter complex deformation of ${\cal N}=4$ SYM, dual to type
IIB string theory in the generalized Lunin-Maldacena backgrounds discovered
by Frolov. We formulate the coordinate space Bethe
ansatz, calculate the $S$-matrix and determine for which choices of parameters
the $S$-matrix fulfills the Yang-Baxter equations. For these 
choices of parameters we furthermore write down the $R$-matrix. 
We find in total four classes of integrable models. In particular,
each already known model of the above type is nothing but one in a family
of such models.}

\keywords{AdS/CFT correspondence, Lunin-Maldacena backgrounds, integrable systems, spin chains, ${\cal N}=4$ SYM}

\begin{document} 


\section{Introduction}
Integrability has played a prominent role in recent years exploration 
of the AdS/CFT correspondence~\cite{Maldacena:1997re}, the starting signal
being the discovery by Minahan and Zarembo that the planar one-loop dilatation
operator of the $\mathfrak{so}(6)$
sector of ${\cal N}=4$ SYM could be identified with
the Hamiltonian of an integrable spin chain~\cite{Minahan:2002ve}.
This discovery was soon extended beyond the $\mathfrak{so}(6)$
sector~\cite{Beisert:2003yb} and beyond the one-loop 
order~\cite{Beisert:2003tq}. On the string theory side integrable structures
first made their appearance with the observation that the Green Schwarz
superstring on $AdS_5\times S^5$ possessed an infinite set of non-local
conserved charges~\cite{Bena:2003wd} and were more precisely described
in a series of important papers~
\cite{Arutyunov:2003uj,Dolan:2003uh,Kazakov:2004qf,Arutyunov:2004yx,Beisert:2005bm,Alday:2005gi}.
An interesting question is to what extent the existence of integrable 
structures in the gauge-string duality is linked to 
supersymmetry.\footnote{It is of course also interesting to study the 
existence of integrable structures for gauge theories for which no string
theory dual is currently known. For a review and further references,
see~\cite{Belitsky:2004cz}.}
 A first
possibility to study this question was provided with the discovery by
Lunin and Maldacena~\cite{Lunin:2005jy} of 
the background required to make type IIB string theory
dual to the Leigh-Strassler $\beta$-deformation of ${\cal N}=4$ 
SYM~\cite{Leigh:1995ep}, 
a conformal quantum field theory carrying ${\cal N}=1$ supersymmetry.
Later the construction of Lunin
and Maldacena was generalized to a three parameter complex deformation of 
${\cal N}=4$ SYM which is still conformal 
(at least to leading order in $N$)
but for which supersymmetry is 
lost~\cite{Frolov:2005dj}. 
Furthermore, it was shown that for uni-modular deformation parameters
one can construct a Lax pair for (the bosonic part of)
the corresponding  classical string theory, showing that this theory like
its undeformed counterpart is integrable~\cite{Frolov:2005dj}. 
(A Lax pair for the Landau-Lifshitz
model describing fast rotating three-spin strings in the deformed 
background~\cite{Frolov:2005iq}
was derived in reference~\cite{Prinsloo:2005dq}).
Let us denote the three deformation parameters
as $r_j e^{i\gamma_j}$,  with $r_j>0$, $\gamma_j$ real, $j\in \{0,1,2\}$.
(For the precise definition, see section~\ref{themodel}.)
The three parameter deformation of 
${\cal N}=4$ SYM reduces to the Leigh-Strassler one when the three
deformation parameters are identical. In~\cite{Beisert:2005if} the general
deformation process was studied from the gauge theory point of view and
it was found that the case $r_0=r_1=r_2=1$ always leads to an integrable
dilatation operator (at least to the order that the latter is known). In 
addition, exact matching of gauge theory anomalous dimensions and string state
energies was found in the two- as well as the three-scalar holomorphic 
sub-sectors (the analogues of the $\mathfrak{su}(2)$
and $\mathfrak{su}(3)$ sub-sectors of the
undeformed theory)~\cite{Frolov:2005iq}.
Results on the existence of integrability on the gauge theory side 
for general deformation parameters  
are mostly negative.
For instance, in the two scalar holomorphic sector integrability holds at 
one loop order
for $r_0=r_1=r_2=r\neq 1$ and
$\gamma_0=\gamma_1=\gamma_2=\gamma$ but is very likely to break down 
already at two-loop order~\cite{Frolov:2005ty}.
Furthermore, it was found in~\cite{Berenstein:2004ys},
see also~\cite{Roiban:2003dw}, that 
assuming $r_0=r_1=r_2=r$ integrability can only be obtained 
at one loop order in the 
three scalar holomorphic sector if $r=1$. 
The one-loop dilatation operator in the three scalar holomorphic sector
of the three parameter complex deformation of the gauge theory is a 
three-state spin chain with nearest neighbour interaction and $U(1)^3$
symmetry. In the present paper we study the most general spin chain
with these properties and determine for which choices of parameters it
can be integrable. We find four classes of integrable models. In particular
each already known integrable model of the above type is nothing but
one in a family of such models. We also rule out 
the possibility that the complex three parameter 
deformation of the one-loop planar
$\mathfrak{su}(3)$ sector of ${\cal N}=4$ could be
integrable under more general circumstances than those already identified.

The organization of our paper is as follows. We begin in section~\ref{themodel}
by writing down the most general Hamiltonian for the type of model we 
wish to consider. Subsequently, in section~\ref{Bethe} we 
determine its $S$-matrix
and in section~\ref{YBEsol} investigate for which choices of parameters this 
$S$-matrix fulfills the Yang-Baxter relation. For these choices of parameters
we then in section~\ref{Rmatrix} 
construct an $R$-matrix for the model. Finally, we compare
our integrable models to those already known in section~\ref{comparison}.
We finish with a brief conclusion.

\section{The model \label{themodel}}

We consider the most general 3-state spin chain with nearest neighbour
interaction and $\mathop{\rm U}(1)^3$ symmetry.
We denote the basis of states at a given site as
$|0\rangle$, $|1\rangle$, and $|2\rangle$. A $\mathop{\rm U}(1)^3$ symmetric Hamiltonian
conserves the number of states of each type and accordingly takes the
form
\begin{eqnarray}\label{eq:Hamiltonian}
\nonumber H&=&H_{00}^{00}E_{00}E_{00}
+H_{11}^{11}E_{11}E_{11}+H_{22}^{22}E_{22}E_{22}+
H_{12}^{12}E_{11}E_{22}+H_{21}^{12}E_{12}E_{21}
\label{Hamiltonian}\\
&& +H_{12}^{21}E_{21}E_{12}+
H_{21}^{21}E_{22}E_{11}+H_{01}^{10}E_{10}E_{01}
+H_{10}^{10}E_{11}E_{00}+H_{01}^{01}E_{00}E_{11}
\nonumber \\
&& +
H_{10}^{01}E_{01}E_{10}+H_{02}^{20}E_{20}E_{02}+H_{20}^{20}E_{22}E_{00}+
H_{02}^{02}E_{00}E_{22}+H_{20}^{02}E_{02}E_{20},
\end{eqnarray}
where $E_{jk}E_{lm}$ is an abbreviation for 
$\sum_{i=1}^L E_{jk}^i E_{lm}^{i+1}$, with $L$ 
being the length of the spin chain, and where
$E_{ij}|k\rangle =|i\rangle \delta_{jk}$, i.e.\
$E_{ij}$ are the generators of $\mathfrak{gl}(3)$. 
The spin chain is assumed to be closed.
Restricting ourselves to Hermitian Hamiltonians we shall assume that
the diagonal elements are real and that the off-diagonal ones can
be written as
\beq\label{gamma}
H_{21}^{12}=(H_{12}^{21})^*\equiv r_0 e^{i \gamma_0},\qquad 
H_{02}^{20}=(H_{20}^{02})^*\equiv r_1 e^{i \gamma_1},\qquad
H_{10}^{01}=(H_{01}^{10})^*\equiv r_2 e^{i \gamma_2},
\eeq
where $r_0$, $r_1$ and $r_2$ are real and positive and where
$\gamma_0$, $\gamma_1$ and $\gamma_2$ are real.  
Our aim is to determine for which values of the parameters this 
Hamiltonian
is integrable. An interesting spin chain which is included in the class of
models given by eqn.~\rf{Hamiltonian} is the spin chain describing the 
 the three scalar holomorphic  
sub-sector of the three parameter complex deformation of ${\cal N}=4$ SYM,
dual to type IIB string theory on 
the generalized Lunin-Maldacena backgrounds found by Frolov.
For this model one has
\bea
&& H_{00}^{00}=H_{11}^{11}=H_{22}^{22}=0,\qquad
H_{12}^{12}=H_{01}^{01}=H_{20}^{20}=1, \nonumber\\
&&H_{21}^{21}=r_0^2,,
\qquad H_{02}^{02}
=r_1^2, \qquad H_{10}^{10}=r_2^2,
\label{complexdeformation} \\ 
&&H_{21}^{12}=(H_{12}^{21})^*\equiv r_0 e^{i \gamma_0},\qquad
H_{10}^{01}=(H_{01}^{10})^*\equiv r_1 e^{i \gamma_1}, \qquad 
H_{02}^{20}=(H_{20}^{02})^*\equiv r_2 e^{i \gamma_2}.\nonumber
\eea
In \cite{Berenstein:2004ys} it was found that (\ref{complexdeformation}) is integrable for $\gamma_0=\gamma_1=\gamma_2=\gamma=\frac{2n\pi}{L}$ with $n$ integer  and
$r_0=r_1=r_2=r$ only if $r=1$. Later the model was shown to be
integrable for $r_0=r_1=r_2=1$ for 
arbitrary values of $\gamma_0$, $\gamma_1$ and 
$\gamma_2$~\cite{Beisert:2005if}.
The case $r_0=r_1=r_2=1$, $\gamma_0=
\gamma_1=\gamma_2=\pi$ is the usual XXX $\mathfrak{su}(3)$ spin chain. 
An additional number of integrable special cases of the model~\rf{Hamiltonian}
are already known. We shall return to these in 
section~\ref{comparison}. The problem of studying the most general 
spin chain with certain global conservation laws has also been pursued
in condensed matter physics, see for instance~\cite{Alcaraz:2003ya}.
Very recently, another problem of a somewhat similar nature was addressed,
namely the problem of determining 
the most general integrable long range spin chain with the
spins transforming in the fundamental of 
$\mathfrak{gl}(n)$~\cite{Beisert:2005wv}.

Integrability implies the existence of an $R$-matrix, $R(u)$,
depending on a spectral parameter $u$ and fulfilling
\beq
R(u)|_{u=u_0}={\cal P},
\hspace{0.9cm} {\cal P}\frac{d}{du}R(u)|_{u=u_0}=H,
\eeq
where ${\cal P}$ is the permutation operator. Furthermore, the $R$-matrix 
must obey  the Yang Baxter equation
\beq
R_{i_1i_2}^{j_1j_2}(u-v)R_{j_1i_3}^{k_1j_3}(u)
R_{j_2j_3}^{k_2k_3}(v)
=R_{i_2i_3}^{j_2j_3}(v)R_{i_1j_3}^{j_1k_3}(u)R_{j_1j_2}^{k_1k_2}(u-v)
\label{YangBaxter}.
\eeq
We can write a Hamiltonian $H$ of the form~\rf{Hamiltonian}
with the angles $\gamma_0$, $\gamma_1$ and $\gamma_2$ present in terms of
a Hamiltonian without angles, $\tilde{H}$,
\begin{equation}
H_{ij}^{kl}=\exp\left(\frac{i}{2}(\epsilon_{ijm}\gamma_m-
\epsilon_{kln}\gamma_n)\right)\tilde{H}_{ij}^{kl}.
\end{equation}
Then, if the Hamiltonian $H$ admits an $R$-matrix, $R_{ij}^{kl}$,
we can construct an $R$-matrix, $\tilde{R}_{ij}^{kl}$, 
corresponding to $\tilde{H}$ by the following
recipe~\cite{Reshetikhin,Beisert:2005if}
\begin{equation}
\tilde{R}_{ij}^{kl}=\exp\left(\frac{i}{2}(\epsilon_{ijm}\gamma_m-\epsilon_{kln}\gamma_n)\right)R_{ij}^{lk}. \label{R}
\end{equation}
Conversely, given an $R$-matrix corresponding to a Hamiltonian of the
form~(\ref{Hamiltonian}) with the angles set to zero, one can construct an
$R$-matrix for a Hamiltonian with the angles re-introduced 
by applying the transformation inverse to~(\ref{R}).
\label{anglediscussion} 
Thus any model integrable without
the presence of angles is also integrable with arbitrary values of
the angles and any
model integrable with non-zero values of the angles
 is also integrable if the angles
are set to zero. In other words, when searching for integrable models,
one can leave out the angles from the analysis.

To test for integrability in a system that admits scattering one can,
instead of directly constructing an $R$-matrix, study the properties of
an appropriately defined $S$-matrix. 
In the present case
the $S$-matrix can be written down by choosing a reference state, say, 
$|0\ldots 0\rangle$ and considering scattering of excitations of type 1 and
2. 
Integrability normally requires non-diffractive and factorized 
scattering.\footnote{
In the case where the angles are present in 
(\ref{gamma}) the 
scattering is actually diffractive, see appendix A.
Due to the argument given above, the model can  
nevertheless  be integrable.} 
This implies that the $S$-matrix must fulfill the Yang-Baxter like
relation
\beq
S_{2,3}(p_2,p_3)S_{1,3}(p_1,p_3)S_{1,2}(p_1,p_2)=
S_{1,2}(p_1,p_2)S_{1,3}(p_1,p_3)S_{2,3}(p_2,p_3),\label{YBS}
\eeq 
where each $S$-matrix acts in a eight-dimensional space.
For a graphical representation of this relation, see Appendix A.
In the following we shall follow the simpler route of first studying the
S-matrix. Hence we start
by considering two particle scattering,  write down the coordinate
space Bethe ansatz, determine the S-matrix and investigate 
for which choices of parameters it fulfills the relation~\rf{YBS}. 
Afterwards, we write down the corresponding $R$-matrix. The coordinate
space Bethe ansatz was introduced by Bethe in 1931~\cite{Bethe:1931hc}
and revived by Staudacher in connection with the study of 
${\cal N}=4$ SYM in~\cite{Staudacher:2004tk}. 
In reference~\cite{Alcaraz:2003ya} a model similar to ours was analyzed 
using the so-called matrix product ansatz, but the analysis was not taken
to the point of actually determining and classifying all integrable cases.

\section{The Bethe Ansatz and the S-matrix \label{Bethe}}

It is obvious that the three states $|0\ldots 0\rangle$, $|1\ldots 1\rangle$
and $|2\ldots 2\rangle$ are all eigenstates of the Hamiltonian. In the
following we shall choose $|0\ldots 0\rangle$ as our reference state.
The integrability properties of the model  
should not depend on
the choice of reference state. First, let us study states containing only
one excitation. We define
\beq
|1\rangle=\sum_{1\leq l_1\leq L}
\psi_{1}(l_1)|00\stackrel{l_1\atop{\downarrow}}{1}000\rangle,
\eeq
and similarly for $|2\rangle$. Writing down the Schr\"{o}dinger equation
gives
\beq
\left(H_{00}^{00}(L-2) +H_{10}^{10}+H_{01}^{01}\right) \psi_1(l_1)
+r_1\left(\psi_1(l_1+1)+\psi_1(l_1-1)\right)=E_1\psi_1(l_1).
\eeq
This equation is immediately solved by the plane wave
\beq
\psi_1(l_1)=e^{ip_1l_1},\hspace{0.7cm} E_1=H_{00}^{00}(L-2)+H_{10}^{10}+H_{01}^{01}+
r_1\left(e^{ip_1}+e^{-ip_1}\right).
\eeq
Similarly, we find a one excitation eigenstate $|2\rangle$ given by
\beq
\psi_2(l_2)=e^{ip_2l_2},\hspace{0.7cm} E_2=H_{00}^{00}(L-2)+H_{20}^{20}+
H_{02}^{02}+
r_2\left(e^{ip_2}+e^{-ip_2}\right).
\eeq
We notice that in general the two types of excitations have different
dispersion relations. Whereas a difference in the $p$-independent terms
is harmless, a difference in the $p$ dependent terms is normally viewed
as a signal that the model is not integrable. We shall make this statement
more precise below. 

Let us now turn to studying  two-body interactions. 
We choose as a basis for the space of two-particle
excitations the states $|11\rangle$, $|12\rangle$, $|21\rangle$ and
$|22\rangle$ where $|ij\rangle$ is defined as
\beq
|ij\rangle=\sum_{1\leq l_1< l_2\leq L}
\psi_{ij}(l_1,l_2)|00\stackrel{l_1\atop{\downarrow}}{i}000
\stackrel{{l_2}\atop{\downarrow}}{j}000\rangle.
\eeq
When $r_0\neq 0$ the model 
allows for scattering between particles of type 1 and 2
and
the states $|12\rangle$ and $|21\rangle$ will mix.
Accordingly, our S-matrix will take the form
\beq\label{S}
S=\pmatrix{a&0&0&0\cr 0&c&b&0\cr
0&\bar{b}&\bar{c}&0\cr 0&0&0&d}.
\eeq
We shall first
consider the case $r_0\neq 0$. Later, we will  comment on the 
simpler case $r_0=0$. 

Let us to begin with study the states $|11\rangle$ and $|22\rangle$ 
which do not
mix with anything. For the former one we get from the Schr\"{o}dinger 
equation
\begin{eqnarray}
\nonumber l_2>l_1+1: && \\ 
 E_{11}\psi_{11}(l_1,l_2)&=&
\left(H_{00}^{00}(L-4)+2H_{10}^{10}+2H_{01}^{01}\right)\psi_{11}(l_1,l_2) \\
&&+r_1\left\{\psi_{11}(l_1+1,l_2)+\psi_{11}(l_1,l_2+1)\right.\nonumber \\
&&\left.+\psi_{11}(l_1-1,l_2)+\psi_{11}(l_1,l_2-1)\right\},\nonumber\\
\nonumber \\
\nonumber l_2=l_1+1:&&\\ 
E_{11}\psi_11(l_1,l_2)&=&
\left(H_{00}^{00}(L-3)+H_{11}^{11}+H_{10}^{10}+H_{01}^{01}\right)\psi_{11}(l_1,l_2)
\\&&+
r_1\left\{\psi_{11}(l_1,l_2+1)+\psi_{11}(l_1-1,l_2)\right\}\nonumber.
\end{eqnarray}
Inserting the standard Bethe ansatz 
\begin{equation}
\psi_{11}(l_1,l_2)=e^{ip_1l_1+ip_2l_2}+a(p_2,p_1)e^{ip_1l_2+ip_2l_1},
\end{equation}
we find 
\beq
E_{11}=H_{00}^{00}(L-4)+2H_{10}^{10}
+2H_{01}^{01}+r_1\left(e^{ip_1}+e^{ip_2}+e^{-ip_1}+e^{-ip_2}\right),
\eeq
and
\beq
a(p_1,p_2)=-\frac{\sigma_1e^{ip_1}+r_1e^{ip_1+ip_2}+r_1}
{\sigma_1e^{ip_2}+r_1e^{ip_1+ip_2}+r_1},
\label{d}
\eeq
where
\beq
\sigma_1=H_{10}^{10}-H_{00}^{00}-H_{11}^{11}+H_{01}^{01}. \label{sigma1}
\eeq
The results from 22-scattering can be obtained from those of
11-scattering by making the replacements $H_{10}^{10}
\rightarrow H_{20}^{20}$,
$H_{01}^{01}\rightarrow H_{02}^{02}$, $r_1\rightarrow r_2$.
Thus we use the Bethe ansatz
\begin{equation}
\psi_{22}(l_1,l_2)=e^{ip_1l_1+ip_2l_2}+d(p_2,p_1)e^{ip_1l_2+ip_2l_1},
\end{equation}
and get
\beq
E_{22}=H_{00}^{00}(L-4)+2H_{20}^{20}
+2H_{02}^{02}+r_2\left(e^{ip_1}+e^{ip_2}+e^{-ip_1}+e^{-ip_2}\right),
\eeq
and
\beq
\label{a}
d(p_1,p_2)=-\frac{\sigma_2e^{ip_1}+r_2e^{ip_1+ip_2}+r_2}
{\sigma_2e^{ip_2}+r_2e^{ip_1+ip_2}+r_2},
\eeq
 where
\beq
\sigma_2=H_{20}^{20}-H_{00}^{00}-H_{22}^{22}+H_{02}^{02}. \label{s2}
\eeq

Next, let us turn to studying the states which mix. For these the
Schr\"{o}dinger equation gives rise to the relations
\begin{eqnarray}
\nonumber l_2>l_1+1:&&\\ \nonumber\\
\label{eq:eom1} E\psi_{12}(l_1,l_2)
&=&
\left(H_{00}^{00}(L-4)+H_{10}^{10}+H_{01}^{01}+
H_{20}^{20}+H_{02}^{02}\right)\psi_{12}(l_1,l_2)\\
&&+r_1\left(\psi_{12}(l_1+1,l_2)+\psi_{12}(l_1-1,l_2)\right)\nonumber \\
&&+r_2\left(\psi_{12}(l_1,l_2+1)+\psi_{12}(l_1,l_2-1)\right),\nonumber \\
\label{eq:eom2} E\psi_{21}(l_1,l_2)&=&
\left(H_{00}^{00}(L-4)+H_{10}^{10}+H_{01}^{01}+H_{20}^{20}+
H_{02}^{02}\right)\psi_{21}(l_1,l_2) \\
&&\nonumber+r_1\left(\psi_{21}(l_1,l_2+1)+\psi_{21}(l_1,l_2-1)\right) \\
&&\nonumber +
r_2\left(\psi_{21}(l_1+1,l_2)+\psi_{21}(l_1-1,l_2)\right),\\
\nonumber \\
\nonumber l_2=l_1+1:&&\\ \nonumber \\
\label{eq:eom3} E\psi_{12}(l_1,l_2)&=&
\left(H_{00}^{00}(L-3)+H_{12}^{12}+H_{01}^{01}+H_{20}^{20}\right)\psi_{12}(l_1,l_2)
\\&&+\nonumber
r_0
\psi_{21}(l_1,l_2)+r_1\psi_{12}(l_1-1,l_2)+r_2\psi_{12}(l_1,l_2+1),\\
\label{eq:eom4} E\psi_{21}(l_1,l_2)&=&\left(H_{00}^{00}(L-3)+
H_{21}^{21}+H_{10}^{10}+H_{02}^{02}\right)\psi_{21}(l_1,l_2)\\
&&+r_0\psi_{12}(l_1,l_2)+
r_1\psi_{21}(l_1,l_2+1)+r_2\psi_{21}(l_1-1,l_2).
\nonumber
\end{eqnarray}
Let us now choose the following Bethe ansatz 
\begin{eqnarray}
\psi_{12}&=&A_{12}e^{ip_1l_1+ip_2l_2}+A_{12}'e^{ip_1'l_2+ip_2'l_1},
\label{psi12}\\
\psi_{21}&=&A_{21}e^{ip_1'l_1+ip_2'l_2}+A_{21}'e^{ip_1l_2+ip_2l_1},
\label{psi21}
\end{eqnarray}
where due to the translational invariance of our model
\beq
p_1+p_2=p_1'+p_2'. \label{momentumcon}
\eeq
The idea is that two excitations with momenta $p_1$ and $p_2$ can scatter 
whereby their momenta get changed to $p_1'$ and $p_2'$. Inserting this 
into (\ref{eq:eom1}) and (\ref{eq:eom2}) we find for the energy
\begin{eqnarray}
\nonumber E&=&H_{00}^{00}(L-4)+H_{10}^{10}
+H_{01}^{01}+H_{20}^{20}+H_{02}^{02}+r_1(e^{ip_1}+e^{-ip_1})+
r_2(e^{ip_2}+e^{-ip_2})\\
&=&H_{00}^{00}(L-4)+H_{10}^{10}+H_{01}^{01}+H_{20}^{20}+H_{02}^{02}+
r_1(e^{ip_2'}+e^{-ip_2'})+r_2(e^{ip_1'}+e^{-ip_1'})
\label{energy}
\end{eqnarray}
Equations~\rf{energy}
and~\rf{momentumcon} determine $p_1'$ and $p_2'$  
\begin{eqnarray}\label{eq:p's}
&& e^{ip_1'}=e^{ip_1}\frac{r_2+r_1e^{ip_1+ip_2}}
{r_1+r_2e^{ip_1+ip_2}},\\
&&e^{ip_2'}=e^{ip_2}\frac{r_1+r_2e^{ip_1+ip_2}}{r_2+r_1e^{ip_1+ip_2}},
\end{eqnarray}
where the solutions are chosen such as to reproduce the standard case
$p_1'=p_1$, $p_2'=p_2$ when $r_1=r_2\neq 0$. If both $r_1$ and $r_2$ 
vanish the usual Bethe ansatz $p_1'=p_1$ and $p_2'=p_2$ is still applicable.
If only one of the two vanishes one necessarily has $p_1'=p_2$ and 
$p_2'=p_1$, i.e. scattering between excitations of  type 1 and 2 is
not possible. In this case one should choose another reference state. 
However, the results for this situation can be obtained by symmetry
arguments from those of
the case $r_0=0$ which we will consider below.

The S-matrix elements involved in the 12-scattering are defined, using the transmission diagonal representation, as
\begin{equation}
\pmatrix{A'_{21}\cr A'_{12}}=
\pmatrix{c(p_2,p_1)&b(p_2,p_1)\cr\bar{b}(p_2,p_1)&
\bar{c}(p_2,p_1)}\pmatrix{A_{12}\cr
A_{21}},
\end{equation}
 and can be found from (\ref{eq:eom3}) and (\ref{eq:eom4}) which
with the Bethe ansatz~\rf{psi12} and~\rf{psi21} read
\bea
0&=&A_{21} r_0 e^{i p_2'}+A_{21}' r_0 e^{i p_1} \\
&&-A_{12}\left\{\tau_1+r_1 e^{i p_1}+r_2 e^{-i p_2}\right\} e^{ip_2} 
-A_{12}'\left\{\tau_1+r_1 e^{ip_2'}+r_2 e^{-ip_1'}\right\}e^{ip_1'}, \nonumber\\
0&=&A_{12} r_0 e^{i p_2}+A_{12}' r_0 e^{i p_1'} \\
&&-A_{21}\left\{\tau_2+r_1 e^{-i p_2'}+r_2 e^{i p_1'}\right\} e^{ip_2'} 
-A_{21}'\left\{\tau_2+r_1 e^{-ip_1}+r_2 e^{ip_2}\right\}e^{ip_1}, \nonumber
\eea
where
\bea
\tau_1&=&H_{10}^{10}-H_{00}^{00}-H_{12}^{12}+H_{02}^{02}, \label{tau1} \\
\tau_2&=&H_{01}^{01}-H_{00}^{00}-H_{21}^{21}+H_{20}^{20}. \label{tau2}
\eea
As mentioned above it is common lore that the model can not be integrable
unless the two scattering excitations have the same dispersion relation.
We will give the precise argument for our Hamiltonian in 
Appendix A.\footnote{Integrable Hamiltonians of the 
type~(\ref{Hamiltonian}) with generic values of the angles, however, constitute
 an exception to the rule, see appendix A.}
>From now on we 
assume that $r_1=r_2$. If in addition $r_0\neq 0$ scattering
will also be possible with the choice of either of the states 
$|1\ldots1\rangle$ or $|2\ldots 2\rangle$ as reference state
and it follows by symmetry arguments that $r_0=r_1=r_2=r$.
For $r_0=r_1=r_2=r\neq 0$ the remaining S-matrix elements 
read

\begin{eqnarray}
\label{eq:c}c(p_1,p_2)&=&\bar{c}(p_1,p_2)=\frac{1}{D}
\left(e^{ip_1}-e^{ip_2}\right)\left(1+e^{ip_1+ip_2}\right),\\
\nonumber b(p_1,p_2)&=&-\frac{1}{D}
\Big((t_1t_2-1)e^{ip_1+ip_2}+(1+e^{ip_1+ip_2})^2\label{b}\\
&&+\left(t_1e^{ip_2}+t_2e^{ip_1}\right)(1+e^{ip_1+ip_2})\Big),\\
\nonumber\bar{b}(p_1,p_2)&=&-\frac{1}{D}\Big((t_1t_2-1)e^{ip_1+ip_2}+(1+e^{ip_1+ip_2})^2
\label{bbar}\\
&&+\left(t_2e^{ip_2}+t_1e^{ip_1}\right)(1+e^{ip_1+ip_2})\Big),
\end{eqnarray}
where 
\begin{equation}
D(p_1,p_2)=\left((t_1t_2-1)e^{2ip_2}+(1+e^{ip_1+ip_2})^2+(t_1+t_2)e^{ip_2}
(1+e^{ip_1+ip_2})\right),\label{D}
\end{equation}
and 
\beq
\label{t} t_1=\tau_1/r, \qquad
t_2=\tau_2/r.
\eeq
It is easy to verify that the S-matrix given by the relations~\rf{d}, 
\rf{a}, \rf{eq:c}, \rf{b} and~\rf{bbar} is unitary.
We notice that effectively the S-matrix (with the restriction 
$r_0=r_1=r_2=r\neq 0$) depends only on the four
parameters  
\beq
s_1=\sigma_1/r, \quad
s_2=\sigma_2/r, \quad
t_1,\quad  t_2, 
\eeq
(cf.\ eqns.~\rf{d}, \rf{a}, 
\rf{eq:c}, \rf{b}, \rf{bbar} and~\rf{D}). 
This is easy to explain. 
To begin with we had 15 parameters, we then removed the angles as integrability properties can be analysed without them and set 
$r_0=r_1=r_2=r$ since the model otherwise is necessarily non-integrable. 
This leaves us with
10 parameters, the single off-diagonal one, $r$, and 9 diagonal ones. First,
one can of course make a global rescaling by $1/r$ without changing the
scattering properties of the model. Secondly, one can construct the following
number operators
\begin{eqnarray}
\hat{N}_0=1\otimes E_{00},&\quad&\hat{M}_0=E_{00}\otimes1,\\
\hat{N}_1=1\otimes E_{11},&\quad&\hat{M}_1=E_{11}\otimes1,\\
\hat{N}_2=1\otimes E_{22},&\quad&\hat{M}_2=E_{22}\otimes1,
\end{eqnarray}
where
$\hat{N}_i$ and $\hat{M}_i$ counts the number of particles of type 
$i$.
Only five of these operators are independent since we have the
relation
\begin{equation}
\hat{N}_0+\hat{N}_1+\hat{N}_2=\hat{M}_0+\hat{M}_1+\hat{M}_2.
\end{equation}
Since the number operators commute with the Hamiltonian, adding such
operators will not change the scattering properties of the 
system.
This means that effectively the S-matrix depends only on four 
parameters.

When $r_1=r_2=0$ we have 
\begin{equation}
c(p_1,p_2)=\bar{c}(p_1,p_2)=0\quad b(p_1,p_2)=\bar{b}(p_1,p_2)=a(p_1,p_2)=d(p_1,p_2)=-e^{ip_1-ip_2}.
\end{equation}

Furthermore, for $r_0=0$, $r_1=r_2=r$ the S-matrix elements involved in
12-scattering read
\bea
c(p_1,p_2)&=&\bar{c}(p_1,p_2)=0, \\
b(p_1,p_2)&=& -\frac{t_2 e^{ip_1}+e^{ip_1+ip_2}+1}{t_2e^{ip_2}+e^{ip_2+ip_1}+1},
\\
\bar{b}(p_1,p_2)&=
&-\frac{t_1 e^{ip_1}+e^{ip_1+ip_2}+1}{t_1e^{ip_2}+e^{ip_2+ip_1}+1}.
\eea

\section{Solution of the YBE's \label{YBEsol}}

A necessary condition for integrability is that the unitary S-matrix
fulfills the Yang-Baxter equation
\beq
S_{2,3}(p_2,p_3)S_{1,3}(p_1,p_3)S_{1,2}(p_1,p_2)=
S_{1,2}(p_1,p_2)S_{1,3}(p_1,p_3)S_{2,3}(p_2,p_3),
\eeq
where
 each S-matrix acts in a 8-dimensional Hilbert space.

For a S-matrix on the form~\rf{S} 
the Yang-Baxter equation gives
rise to 7 relations between  matrix elements,
\begin{eqnarray}
&&\label{eq:YBE1}\bar{b}b'\bar{b}''-b\bar{b}'b''=0,\\
&&\label{eq:YBE2}a\bar{b}'a''-\bar{b}a'\bar{b}''-\bar{c}\bar{b}'c''=0,\\
&&\label{eq:YBE3}ca'b''-ac'b''+\bar{b}b'c''=0,\\
&&\label{eq:YBE4}\bar{b}\bar{c}'a''-\bar{c}\bar{b}'b''-\bar{b}a'\bar{c}''=0,\\
&&\label{eq:YBE5}d\bar{b}'d''-\bar{b}d'\bar{b}''-c\bar{b}'\bar{c}''=0,\\
&&\label{eq:YBE6}\bar{c}d'\bar{b}''-d\bar{c}'\bar{b}''+b\bar{b}'\bar{c}''=0,\\
&&\label{eq:YBE7}\bar{b}c'd''-c\bar{b}'b''-\bar{b}d'c''=0,
\end{eqnarray}
where $b=b(p_1,p_2)$, $b'=b(p_1,p_3)$, $b''=b(p_2,p_3)$, etc.
At first sight these equations seem rather involved but a systematic
investigation with the aid of Mathematica allows one to find the most
general solution by purely {\it analytical} means. The simplest equation
is equation~\rf{eq:YBE1} so this is the one to address first.
Next, we note that for our case, where $c=\bar{c}$, 
equations~\rf{eq:YBE5},~\rf{eq:YBE6} 
and~\rf{eq:YBE7} appear from equations~\rf{eq:YBE2},~\rf{eq:YBE3}
and~\rf{eq:YBE4} by the interchangement $a\rightarrow d$. Hence
the solution of the former three equations can immediately be read 
off from the solution of the latter three.

\paragraph{The case $r_0=r_1=r_2=r\neq0$:} 

In this situation the equation~\rf{eq:YBE1} gives that $t_2 t_1=1$ and the
equations~\rf{eq:YBE2}, \rf{eq:YBE3} and~\rf{eq:YBE4} give $s_1=0$ or
$s_1=t_1+t_2$. Thus we find the following four families of integrable
models 
\begin{enumerate}
\item
$t_2t_1=1$,\,\, $s_1=s_2=0.$
\item
$t_2t_1=1$,\,\, $s_1=0$,\,\, $s_2=t_1+t_2.$
\item
$t_2t_1=1$, \,\,$s_1=t_1+t_2$,\,\, $s_2=0.$
\item
$t_2t_1=1$,\,\, $s_1=s_2=t_1+t_2$.
\end{enumerate}
It is straightforward to check that these criteria for integrability do not
depend on the choice of reference state. More precisely, the first three
cases are related by $Z_3$ symmetry, i.e.\ they appear from one another
when the labels 0, 1 and 2 are interchanged. The last is invariant under this
symmetry. Thus, we have only two genuinely different classes of integrable
models with $r_1=r_2=r_3\neq 0$.

\paragraph{The case $r_0\neq 0$, $r_1=r_2=0$:} In this case it is obvious
that the YBE's are always fulfilled since all the S-matrix elements
are identical.

\paragraph{The case $r_0=0$, $r_1=r_2=r\neq0$:}
Here the condition for integrability reads
$$
t_1=t_2=s_1=s_2.
$$
Finally, the model is obviously integrable for $r_0=r_1=r_2=0$.

\section{R-matrices \label{Rmatrix}}

In this section we write down the $R$-matrices corresponding to our
integrable Hamiltonians with $r_1=r_2=r_3\neq0$.
These Hamiltonians all have their
off-diagonal elements equal to one. They can therefore be characterized
entirely in terms of their diagonal elements. We can choose representatives
for the four different cases above for instance
as follows
\begin{enumerate}
\item
$H_1:(0,\frac{1}{t},\frac{1}{t},t,
t+\frac{1}{t},\frac{1}{t},t,t,t+\frac{1}{t}),$
\item
$H_2:(t+\frac{1}{t},\frac{1}{t},\frac{1}{t},t,0,\frac{1}{t},t,t,t+\frac{1}{t}),$
\item
$H_3:(t+\frac{1}{t},\frac{1}{t},\frac{1}{t},t,t+\frac{1}{t},\frac{1}{t},t,t,0),$
\item
$H_4:(t+\frac{1}{t},\frac{1}{t},\frac{1}{t},t,t+\frac{1}{t},\frac{1}{t},t,t,
t+\frac{1}{t})$,
\end{enumerate}
where the lists are lists of diagonal elements. Any member of a given class
can be brought on the form above by addition of appropriate linear combinations
of number operators. 
We can write the $R$-matrix for the models above in a collective form as
$$
R_{00}^{00}(u,v)=A(u,v,s'),\,\,\,\,\,
R_{11}^{11}(u,v)=A(u,v,\tilde{s}),\,\,\,\,\,
R_{22}^{22}(u,v)=A(u,v,s), $$
$$R_{12}^{12}(u,v)=R_{21}^{21}(u,v)=
R_{02}^{02}(u,v)=R_{20}^{20}(u,v)=
R_{01}^{01}(u,v)=R_{10}^{10}(u,v)=C(u,v),$$
$$R_{01}^{10}(u,v)= R_{02}^{20}(u,v)=R_{12}^{21}(u,v)=B(u,v,t),$$
$$R_{10}^{01}(u,v)= R_{20}^{02}(u,v)=R_{21}^{12}(u,v)=B(u,v,\frac{1}{t}),$$
where
\bea
A(u,v,s)&=&\frac{s u+uv+1}{sv+uv+1}, \\
B(u,v,t)&=&\frac{1+uv+\frac{1}{t}u+tv}{1+u v+(\frac{1}{t}+t)v},\\
C(u,v)&=&\frac{u-v}{1+uv+(t+\frac{1}{t})v},  
\eea
and where the remaining $R$-matrix elements vanish. 
The appropriate choices of $s$, $\tilde{s}$ and $s'$ 
for the four Hamiltonians above
are
\begin{enumerate}
\item
$H_1: $ \,\,\, $(s,s',\tilde{s})
=(t+\frac{1}{t},0,t+\frac{1}{t}),$
\item
$H_2:$ \,\,\, $(s,s',\tilde{s})
=(t+\frac{1}{t},t+\frac{1}{t},0),$
\item
$H_3: $ \,\,\, $(s,s',\tilde{s})
=(0,t+\frac{1}{t},t+\frac{1}{t}),$
\item
$H_4: $ \,\,\, $(s,s',\tilde{s})
=(t+\frac{1}{t},t+\frac{1}{t},t+\frac{1}{t})$.
\end{enumerate}
The R-matrix fulfills the necessary requirements, namely
\beq
R(u,u)=P,
\eeq
where $P$ is the permutation operator and
\begin{equation}
\left(1+v^2+\left(t+\frac{1}{t}\right)v\right)^{-1}H=P\partial_u R(u,v)|_{u=v}.
\end{equation}
Finally, it also satisfies the Yang Baxter relation
\beq
R_{i_1i_2}^{j_1j_2}(u,v)R_{j_1i_3}^{k_1j_3}(u,w)
R_{j_2j_3}^{k_2k_3}(v,w)
=R_{i_2i_3}^{j_2j_3}(v,w)R_{i_1j_3}^{j_1k_3}(u,w)R_{j_1j_2}^{k_1k_2}(u,v).
\eeq
As indicated by the notation above the $R$-matrix has been constructed from
pieces from the earlier determined $S$-matrix, cf.\  eqns.~\rf{S},
\rf{a}, \rf{d}, \rf{eq:c}, \rf{b} and~\rf{bbar}.
In many situations one can perform a transformation from $(u,v)$ 
to a new set of variables $(\lambda,\nu)$
such that  $R(\lambda,\nu)=R(\lambda-\nu)$. 
For instance, for the $\mathfrak{su}(3)$ XXX spin chain, which belongs to case four
and has $t=1$, this transformation
reads
\beq
u=-\frac{\lambda-i}{\lambda+i},\hspace{0.7cm}v= -\frac{\nu-i}{\nu+i}.
\eeq

\section{Comparison to known models \label{comparison}}

Here we list a number of already known integrable models whose Hamiltonian
can be written on the form~\rf{Hamiltonian}.

\paragraph{}

The first example of an integrable model of the form~\rf{Hamiltonian} which 
comes to mind is the XXX $\mathfrak{su}(3)$ spin chain. It is characterized by
the elements of the Hamiltonian taking the  values
$H_{00}^{00}=H_{11}^{11}=H_{22}^{22}=0$,
$H_{12}^{12}=H_{21}^{21}=H_{10}^{10}=H_{01}^{01}=H_{20}^{20}=
H_{02}^{02}=1$, $H_{21}^{12}=H_{12}^{21}=
H_{01}^{10}=H_{10}^{01}=H_{02}^{20}=H_{20}^{02}=-1$
or equivalently
$r=1$, $s_1=s_2=2$, $t_1=t_2=1$. (Note that in the latter notation the angles
are removed from the analysis.)
It thus belongs to family number four.
\paragraph{}

In the same family we find the integrable deformations of this spin chain, 
describing the dilatation operator of
the three scalar holomorphic sub-sector 
of the three parameter complex deformation of ${\cal N}=4$ SYM, given
by~\rf{complexdeformation} with 
$r_0=r_1=r_2=1$~\cite{Berenstein:2004ys,Beisert:2005if}. 
They all have  $s_1=s_2=2$, $t_1=t_2=1$. 

Within our formalism we can of course investigate whether it is possible to
achieve integrability  for a more general class of complex deformations
of ${\cal N}=4$ SYM. 
The most interesting case is the case where all three $r$-variables are
non-vanishing. First, we have seen that integrability demands that
$r_0=r_1=r_2=r$. In reference~\cite{Berenstein:2004ys} it was argued that 
if $r_0=r_1=r_2=r$ one  furthermore needs that $r=1$. 
This also follows from the results above.
Namely, for the deformed model with $r_0=r_1=r_2=r$ we have
\bea
s_1&=&s_2=\frac{1+r^2}{r}, \\
t_1&=&\frac{2r^2-1}{r}, \\
t_2&=&\frac{2-r^2}{r},
\eea
and we immediately see that according to the conditions for integrability
presented above the model can only be integrable if $r=1$ since only
in this case $t_1t_2=1$.
\paragraph{}
In family four we also find the integrable $\mathfrak{su_q}(3)$ spin chain 
likewise studied in~\cite{Berenstein:2004ys} for its possible connection
to other deformations of ${\cal N}=4$ SYM. This model is characterized by
 $H_{00}^{00}=H_{11}^{11}=H_{22}^{22}=0$,
$H_{12}^{12}=H_{01}^{01}=H_{02}^{02}=1$, 
$H_{21}^{12}=H_{12}^{21}=H_{01}^{10}=H_{10}^{01}=H_{02}^{20}=H_{20}^{02}=r$,
$H_{21}^{21}=H_{10}^{10}=H_{20}^{20}=r^2$ or equivalently
$s_1=s_2=\frac{1+r^2}{r}$, $t_1=r$,
$t_2=1/r$.
\paragraph{}
In family three we find f.\ inst.\ the $\mathfrak{su}(1|2)$ spin chain 
describing the $\mathfrak{su}(1|2)$ sub-sector of undeformed 
${\cal N}=4$ SYM and studied 
in~\cite{Beisert:2005fw}. Here one has $H_{00}^{00}=0, H_{11}^{11}=2$, $H_{22}^{22}=0$,
$H_{12}^{12}=H_{21}^{12}=H_{12}^{21}=H_{21}^{21}=H_{10}^{10}=H_{01}^{01}=H_{20}^{20}=
H_{02}^{02}=1$, $H_{01}^{10}=H_{10}^{01}=H_{02}^{20}=H_{20}^{02}=-1$ 
or equivalently $r=1$, $s_1=0$, $s_2=2$,
$t_1=t_2=1$.
 The spin chain describing the analogue of the 
$\mathfrak{su}(1|2)$ sub-sector in the three parameter 
complex deformation of ${\cal N}=4$ SYM 
theory is of course also included here. It has 
$s_1=r+\frac{1}{r}-2$, $s_2=r+\frac{1}{r}$, $t_1=2r-\frac{1}{r}$, 
$t_2=\frac{2}{r}-r $ and is only integrable if $r=1$ where it reduces
to the model considered in~\cite{Beisert:2005if}.
\paragraph{}
We finally note some models known from studies of integrability in the context of condensed matter systems, see for instance \cite{Alcaraz:2003ya}. Choosing the parameters as $H_{00}^{00}=\epsilon_1\cosh\gamma$, $H_{11}^{11}=\epsilon_2\cosh\gamma$, $H_{22}^{22}=\epsilon_3\cosh\gamma$, $H_{21}^{21}=H_{10}^{10}=H_{20}^{20}=
-H_{12}^{12}=-H_{01}^{01}=-H_{02}^{02}=\sinh\gamma$ and $H_{21}^{12}=H_{12}^{21}=H_{01}^{10}=H_{10}^{01}=H_{02}^{20}=H_{20}^{02}=1$. Choosing $\epsilon_1=\epsilon_2=\epsilon_3=\pm1$ this is the anisotropic Perk-Schultz model \cite{Perk:1981nb} which then belongs to class 4. Setting $\gamma=0$ in this model we find the $\mathfrak{su}(3)$ Sutherland model \cite{Sutherland:1975vr}. Another example is the case when $\epsilon_1=-\epsilon_2=\epsilon_3=1$ which is the anisotropic supersymmetric t-J model \cite{schlo}, this model belongs to our class 2.

\section{Conclusion {\label{Conclusion}} }

We have studied the most general three-state spin chain with nearest neighbour
interaction and $U(1)^3$ symmetry. This model contains as a special case the
spin chain describing the three scalar holomorphic sub-sector of the three
parameter complex deformation of ${\cal N}=4$ SYM, dual to type IIB string
theory in the generalized Lunin-Maldacena backgrounds found by 
Frolov~\cite{Frolov:2005dj}.
We have made use of the conceptually simple coordinate space Bethe ansatz
invented by Bethe in 1931~\cite{Bethe:1931hc} and revived in 
connection with the
study of ${\cal N}=4$ SYM by Staudacher~\cite{Staudacher:2004tk}. Solving the
Yang Baxter relation for the $S$-matrix we identified four classes of
integrable models. Subsequently we wrote
down an $R$-matrix for the most interesting of these.
Our findings show that each already known integrable 
model of the above type is nothing but one in a family of such models.
We furthermore rule out 
the possibility that the complex three parameter 
deformation of the one-loop planar
$\mathfrak{su}(3)$ sector of ${\cal N}=4$ could be
integrable under more general circumstances than those already identified.

\acknowledgments

It is a pleasure to thank Niklas Beisert, Vadim Cheianov, 
Peter Orland, Konstantin Zarembo 
and Mikhail Zvonarev for useful discussions. 
C.K.\ acknowledges the support of 
ENRAGE (European Network on Random Geometry), a Marie Curie
Research Training Network supported by the European Community's
Sixth Framework Programme, network contract MRTN-CT-2004-005616.

\appendix
\section{The case of different dispersion relations}
The Yang Baxter relation~\rf{YBS} 
which describes the scattering between three
excitatons has the well-known graphical interpretation 
shown in figure~\ref{YBfigure}. A necessary condition for the relation to be
fulfilled is that incoming and outgoing momenta are the same on
the two sides of the equality sign. It is easy to find an example
where this is  not the case if $r_2\neq r_1$. For $r_0\neq0$ one
can f.\ inst.\ consider
the situation shown in figure~\ref{112211}. Here the incoming particles are
supposed to be of type 1, 1 and 2 with momenta $q_1$, $q_2$ and $q_3$.
The outgoing particles are assumed to be of type 2, 1 and 1 and 
their momenta can be found by the use of~(\ref{eq:p's}). It is easy to
see that 
the outgoing momenta on the two sides of the equation do not match
unless $r_2=r_1$. Since the integrability properties of the model
can not depend on the choice of reference state the model can thus only
be integrable if $r_0=r_1=r_2$. For $r_0=0$ excitations of type
1 and 2 can not cross each but we can now in stead consider the diagram
shown
in figure~\ref{211211}. Again, we reach the conclusion that $r_2=r_1$.

\FIGURE[ht]{\label{YBfigure}
{$\sum_{q'_1,q'_2,q'_3}$}
\parbox{4.5cm}{\centering\includegraphics[height=4cm]{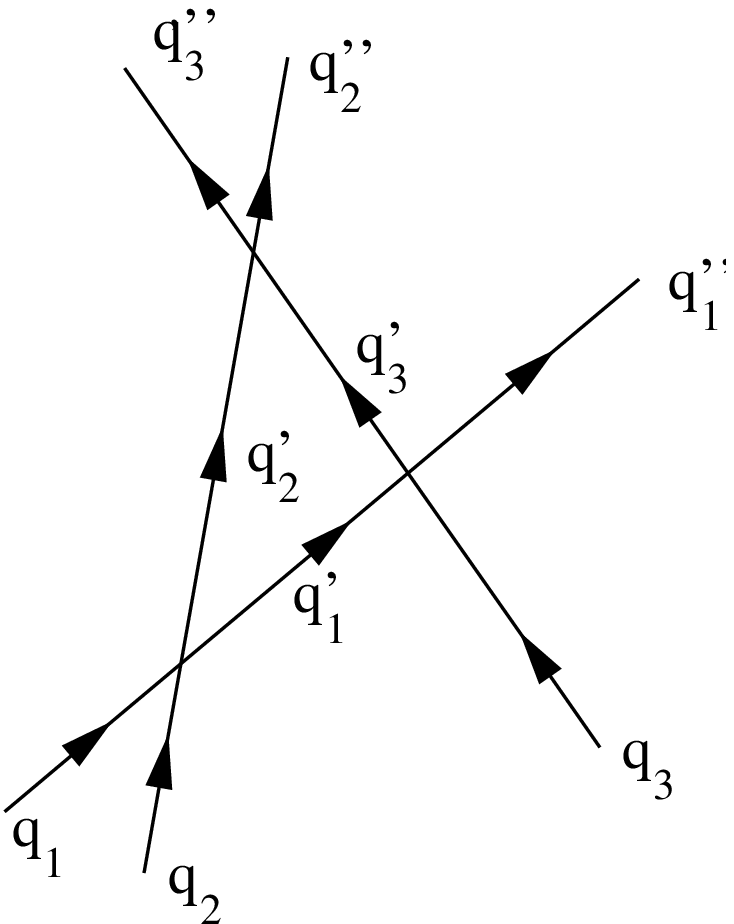}}
$=\quad$
{$\sum_{q'_1,q'_2,q'_3}$}
\parbox{4.5cm}{\centering\includegraphics[height=4cm]{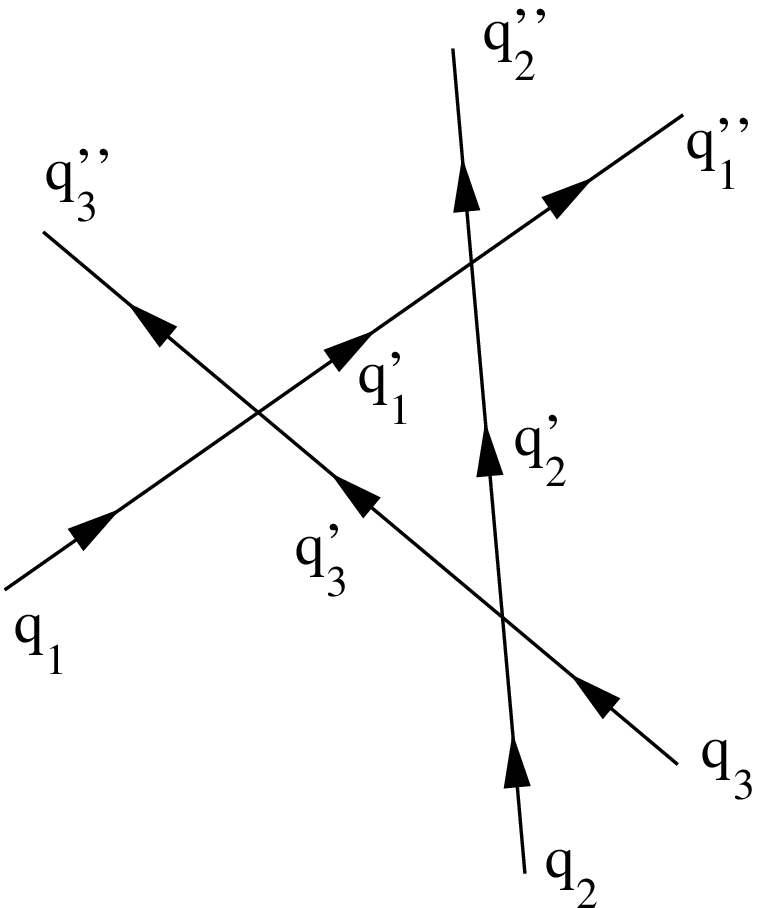}}
\caption{Graphical representation of the Yang-Baxter relations}
}

\FIGURE[ht]{\label{112211}
\parbox{4.5cm}{\centering\includegraphics[height=4cm]{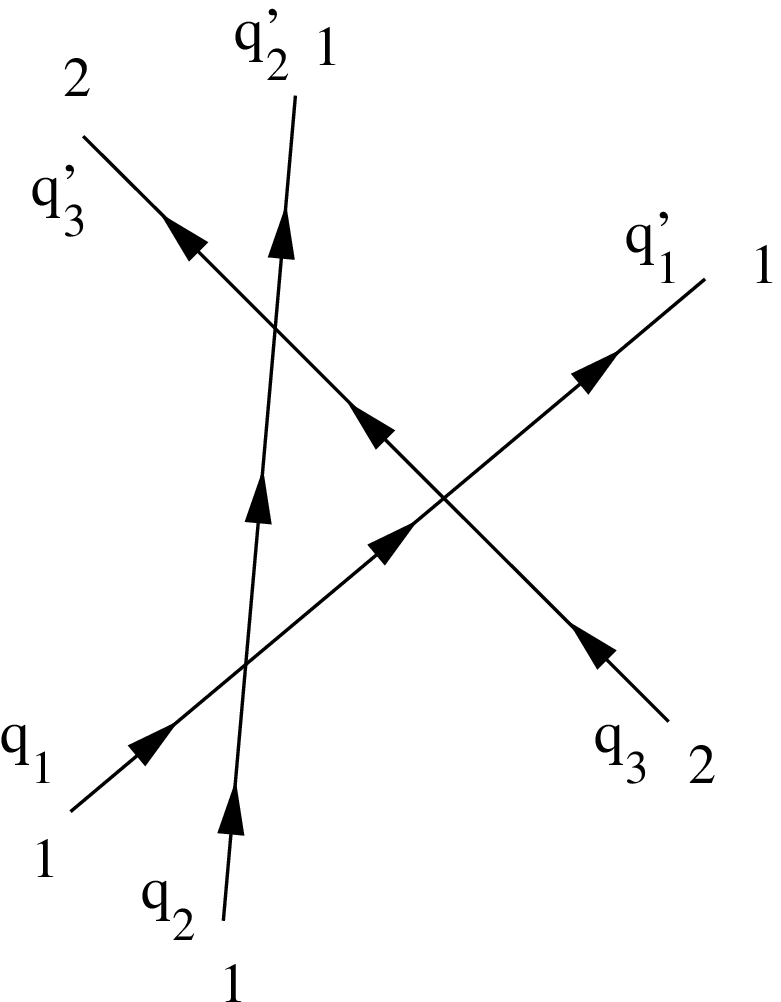}}
=
\parbox{4.5cm}{\centering\includegraphics[height=4cm]{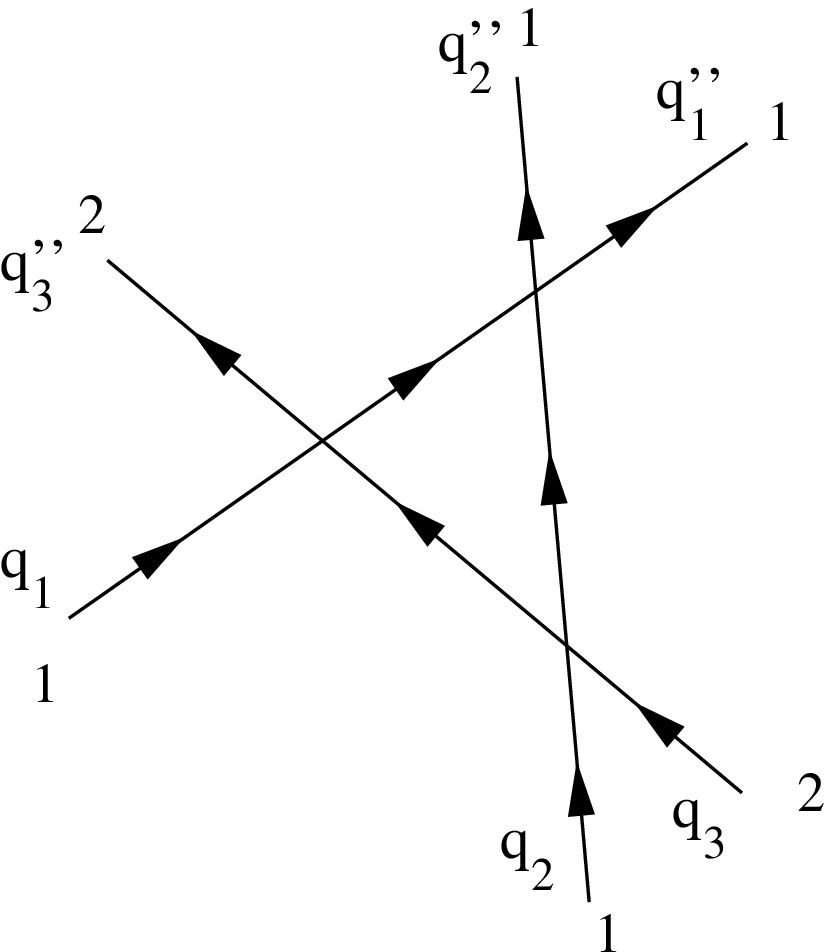}}
\caption{Factorized scattering requires this relation to hold. This is not possible unless the incoming and  outgoing momenta are the same on both sides of the equality sign. Requiring that $q_i'=q_i''$ implies that $r_2=r_1$.}
}
\FIGURE[ht]{\label{211211}
\parbox{4.5cm}{\centering\includegraphics[height=4cm]{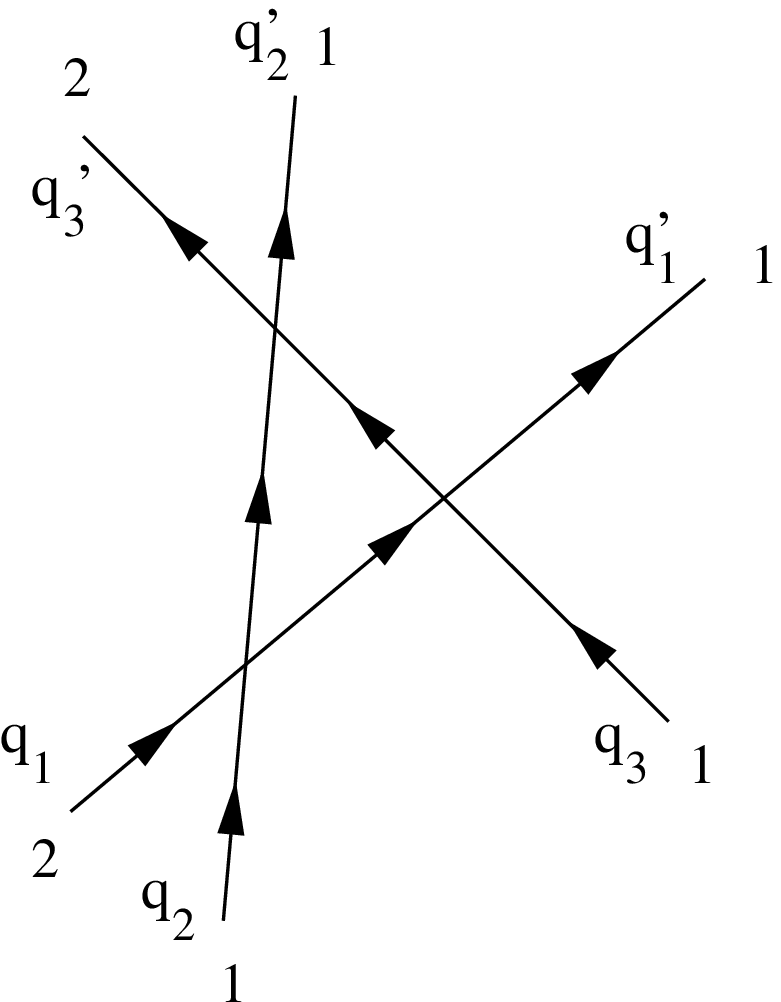}}
=
\parbox{4.5cm}{\centering\includegraphics[height=4cm]{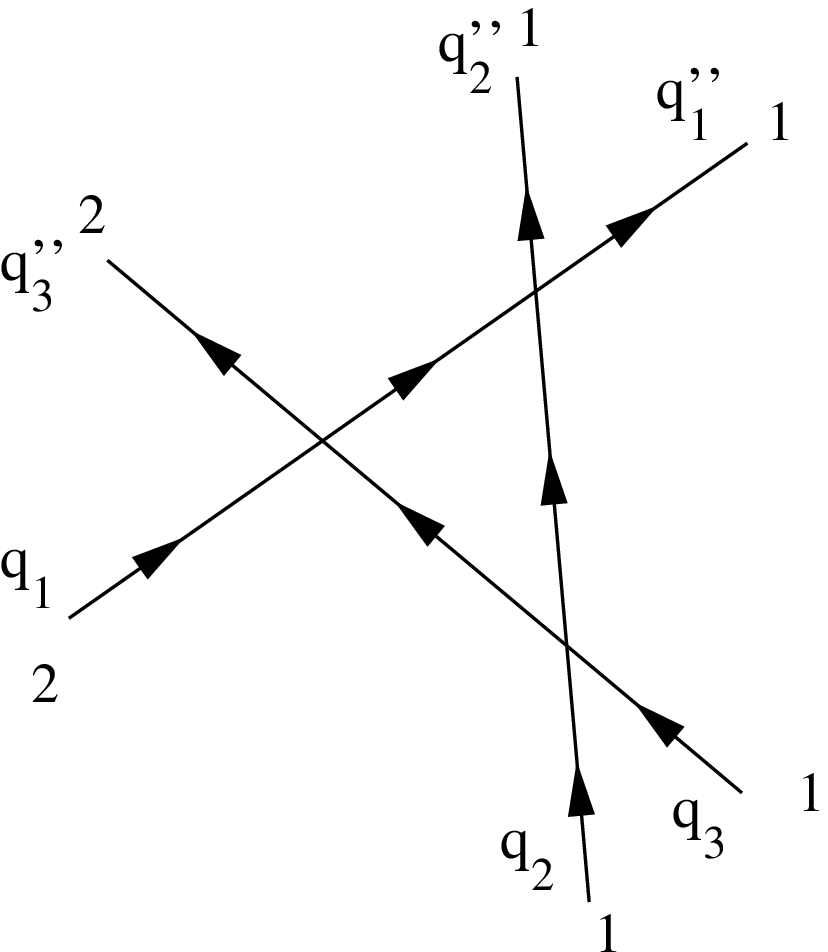}}
\caption{In the case $r_0=0$ this process determines that $r_1=r_2$.}
}
Let us finally comment on the case where the angles are present. In that case the analogue of (\ref{eq:p's}) reads:
\begin{eqnarray}
e^{ip_1'}&=&e^{ip_1}e^{-i(\gamma_1+\gamma_2)}\frac{r_2+r_1e^{i(\gamma_2-\gamma_1)}e^{ip_1+ip_2}}{r_2+r_1e^{2i(\gamma_2-\gamma_1)}e^{ip_1+ip_2}},\\
e^{ip_2'}&=&e^{ip_2}e^{i(\gamma_2+\gamma_1)}\frac{r_1+r_2e^{i(\gamma_2-\gamma_1)}e^{ip_1+ip_2}}{r_1+r_2e^{2i(\gamma_2-\gamma_1)}e^{ip_1+ip_2}}.
\end{eqnarray}
Setting $r_2=r_1$ we find
\begin{equation}
p_1'=p_1-(\gamma_2+\gamma_1),\qquad p_2'=p_2+(\gamma_2+\gamma_1).
\end{equation}
This means that the outgoing momenta in for instance figure 2 are
\begin{equation}
q_1'=q_1''=q_1-(\gamma_2+\gamma_1),\qquad 
q_2'=q_2''=q_2-(\gamma_2+\gamma_1),\qquad q_3'=q_3''=q_3+2(\gamma_2+\gamma_1).
\end{equation}
The outgoing momenta are hence the same on the two sides of the equation but
the scattering is diffractive. However, if the system is integrable with
the angles set to zero it is still integrable when the angles are introduced,
cf.\ discussion on page~\pageref{anglediscussion}.

\end{document}